\documentclass[aps,pre,twocolumn,superscriptaddress,showpacs,floatfix]{revtex4}
\usepackage{amssymb}
\usepackage{graphicx}
\begin{document}

% Use the \preprint command to place your local institutional report
% number in the upper righthand corner of the title page in preprint mode.
% Multiple \preprint commands are allowed.
% Use the 'preprintnumbers' class option to override journal defaults
% to display numbers if necessary
%\preprint{}

%Title of paper
\title{Quasiperiodic localized oscillating 
solutions in the discrete nonlinear 
Schr{\"o}dinger equation with alternating on-site potential}

% repeat the \author .. \affiliation  etc. as needed
% \email, \thanks, \homepage, \altaffiliation all apply to the current
% author. Explanatory text should go in the []'s, actual e-mail
% address or url should go in the {}'s for \email and \homepage.
% Please use the appropriate macro foreach each type of information

% \affiliation command applies to all authors since the last
% \affiliation command. The \affiliation command should follow the
% other information
% \affiliation can be followed by \email, \homepage, \thanks as well.
\author{Magnus Johansson}
\email[]{mjn@ifm.liu.se}
\homepage[]{http://www.ifm.liu.se/~majoh}
%\thanks{}
%\altaffiliation{}
\affiliation{Dept.\ of Physics and Measurement Technology, 
Link\"oping University, S-581 83 Link\"oping, Sweden}

\author{Andrey V. Gorbach}
\email[]{gorbach@mpipks-dresden.mpg.de}
%\homepage[]{http://www.ifm.liu.se/~majoh}
%\thanks{}
%\altaffiliation{}
\affiliation{Max-Planck-Institut f{\"u}r Physik komplexer Systeme,
N{\"o}thnitzer Str.\ 38,
01187 Dresden, Germany}

%Collaboration name if desired (requires use of superscriptaddress
%option in \documentclass). \noaffiliation is required (may also be
%used with the \author command).
%\collaboration can be followed by \email, \homepage, \thanks as well.
%\collaboration{}
%\noaffiliation

\date{\today}

\begin{abstract}
We present what we believe to be the first known example of an exact 
quasiperiodic localized stable solution with spatially symmetric 
large-amplitude oscillations in a non-integrable 
Hamiltonian lattice model. The model is a one-dimensional discrete nonlinear 
Schr{\"o}dinger equation with alternating on-site energies, modelling e.g. 
an array of optical waveguides with 
alternating widths. The solution bifurcates 
from a stationary discrete gap 
soliton, and in a regime of large oscillations its intensity oscillates 
periodically between having one peak at the central site, and two symmetric 
peaks at the neighboring sites with a dip in the middle. Such solutions, 
termed 'pulsons',  
are found to exist in continuous families ranging arbitrarily close both 
to the anticontinuous and continuous limits. Furthermore, it is shown that 
they may be linearly stable also in a regime of large oscillations. 

% insert abstract here
\end{abstract}

% insert suggested PACS numbers in braces on next line
% insert suggested keywords - APS authors don't need to do this
%\keywords{}

%\maketitle must follow title, authors, abstract, \pacs, and \keywords
\maketitle

% body of paper here - Use proper section commands
% References should be done using the \cite, \ref, and \label commands
%\section{}
% Put \label in argument of \section for cross-referencing
%\section{\label{}}
%\subsection{}
%\subsubsection{}

The discrete nonlinear Schr{\"o}dinger (DNLS) equation is one of the most 
studied examples of a non-integrable Hamiltonian lattice model. It is of 
great interest as well from a general nonlinear dynamics point of view, where 
it provides a particularly simple system to analyze fundamental phenomena 
arising from competition of nonlinearity and discreteness such as energy 
localization, wave instabilities etc., as from a more applied viewpoint 
describing e.g.\ arrays of nonlinear optical waveguides \cite{waveguides} 
or Bose-Einstein 
condensates in external periodic potentials \cite{BEC}. 
For recent reviews of the 
history, properties and applications of DNLS-like models, 
see Refs.\ \cite{KRB01,EJ}.

Recent attention has been given to DNLS-like models having, in 
addition 
to the fundamental periodicity given by the lattice constant, also a 
superlattice modulation creating a second period. In particular, the DNLS 
equation with an 
additional term corresponding to alternating on-site energies has been 
proposed to model an optical waveguide array where the individual 
wave\-guides 
have alternating widths \cite{SK}. Creating 
a two-branch linear dispersion relation, the superperiodicity thus provides 
a possibility for existence of 
a new type of nonlinear localized modes, {\em discrete gap solitons} (or 
{\em discrete gap breathers}), with frequencies in the gap between the upper 
and lower branches. These modes appear as {\em stationary} 
solutions to the DNLS 
equation (i.e., with 
constant intensity and a purely harmonic time-dependence 
which can be removed 
by transforming into a rotating frame), and their properties 
were recently analyzed in detail 
in Ref.\ \cite{GJ03}, to which we also refer for further references on 
the topic. Very recently, they have also been experimentally 
observed \cite{gapexp}.

However, it is known that as the DNLS equation 
in one aspect is non-generic among nonlinear Hamiltonian lattice models, 
with a second conserved quantity being the total norm of the excitation, 
there exist also localized {\em quasiperiodic} solutions which may have 
two (generally) incommensurate frequencies. Here the first frequency 
corresponds to harmonic oscillations at constant intensity as for 
stationary solutions, while the second frequency corresponds to 
time-periodic 
{\em oscillations of the intensity} in the frame rotating with the first 
frequency. The existence of such solutions was proposed in 
Ref.\ \cite{MacKay}, and later explicit examples were constructed and 
analyzed by continuation of multi-site breathers from the 
'{\em anticontinuous}' limit of zero intersite coupling \cite{JA97} 
(note that similar ideas were used already in  
Ref.\ \cite{Nussbaum}  for finite systems). A 
rigorous approach to the connection between existence of quasiperiodic 
solutions and additional conserved quantities was given in 
Ref.\ \cite{BV02}. 
A slightly different approach was taken in Ref.\ \cite{KW03} (see also 
\cite{KRB01}), where the existence of exact quasiperiodic solutions 
bifurcating from localized eigenmodes of the {\em linearized} equations of 
motion around some particular stationary solutions was shown. In all cases, 
to guarantee localization it is necessary that a {\em non-resonance} 
condition is fulfilled, so that no higher harmonics enter the continuous 
linear spectrum. 

However, all known examples of stable 
localized exact quasiperiodic solutions to the 
ordinary DNLS 
equation could be considered as rather special, since (i) they only exist 
in bounded parameter intervals at weak inter-site coupling, and (ii) 
the intensity oscillations are typically quite small compared to their 
average values. Furthermore, to the best of our knowledge no explicit 
example of a stable quasiperiodic breather with {\em spatial symmetry} 
has been given (although one of the modes presented in 
Ref.\ \cite{KW03} possibly 
could yield such a solution). On the other hand, it was shown already in 
Ref.\ \cite{Mez} that for the two-dimensional DNLS model a state with 
large-amplitude symmetric intensity 
oscillations, with the intensity maximum periodically oscillating between 
the central site and four surrounding sites, was created in an intermediate 
stage in the process of 'quasicollapse' of a broad excitation to a highly 
localized breather. This entity, termed '{\em pulson}', 
typically disappeared 
after 3-4 oscillations, transforming into an on-site localized stationary 
breather with slowly decaying small-amplitude internal mode oscillations 
(the decay of which can be calculated similarly as for the one-dimensional 
case in Ref.\ \cite{growth}). 

It is the purpose of the present Brief Report 
to provide the first example of an {\em exact} stable solution with such 
pulson properties, which we have found to exist in the one-dimensional 
binary modulated DNLS equation. In fact, we remarked already in 
Ref.\ \cite{GJ03} that similar arguments as in Ref.\ \cite{KW03} prove the 
existence of exact quasiperiodic solutions bifurcating from certain internal 
modes of the stationary gap breathers; here the crucial property (which is 
not fulfilled e.g.\ for single-site breathers in the nonmodulated DNLS 
model \cite{growth}) is, that these internal modes have frequencies 
{\em above} the continuous spectrum, and consequently also all higher 
harmonics will lie outside the continuous spectrum. Referring to the 
notation in Fig.\ 3 of Ref.\ \cite{GJ03}, we here concentrate on the 
spatially symmetric mode denoted as 'P1', which may exist arbitrarily close 
to as well the anticontinuous as the continuous limit, and show that the 
families of quasiperiodic solutions bifurcating from this mode indeed 
exhibit pulson properties as the oscillation amplitude increases. 

Using for convenience a slightly different notation than in 
Ref.\ \cite{GJ03}, we consider the DNLS equation in the form:
\begin{equation}
 i \dot{\psi}_{n}-V_0 (-1)^n \psi_n + C (\psi_{n+1}+\psi_{n-1}) 
+ |\psi_{n}|^{2}\psi_{n} = 0 ,
 	\label{DNLS}
 \end{equation}
with the two conserved quantities
 Hamiltonian 
$
H = 
\sum_n \left [ V_0 (-1)^n |\psi_n|^2 - 
C ( \psi_n \psi_{n+1}^\ast + \psi_n^\ast \psi_{n+1} ) -
\frac{1}{2}| \psi_n|^4  \right ] ,
$
and norm ${\cal N}= \sum_n |\psi_n|^2 $.
Writing a stationary solution as 
$\psi_n (t) = \phi_n^{(\Lambda)} e^{i \Lambda t}$ with 
time-independent $\phi_n^{(\Lambda)}$, 
the linear dispersion relation becomes 
$\Lambda= \pm \sqrt{V_0^2 + 4 C^2 \cos^2 q}$ with gap 
$\Lambda \in [-|V_0|, |V_0|]$. Assuming $V_0 >0$, a discrete gap soliton 
then bifurcates from the {\em lower} gap edge $\Lambda=-V_0$, and 
corresponds in the limit $C\rightarrow 0$ to a single excited {\em even} 
site with intensity $|\psi_n|^2 = \Lambda + V_0$. 

Exact quasiperiodic solutions with two independent frequencies, 
$\omega_0$ and $\omega_b$, can then be found numerically to computer 
precision by using essentially 
the same method as in 
Ref.\ \cite{JA97}. First, we transform into a frame rotating with the  
frequency $\omega_0$, $\psi_n(t) = \phi_n(t)  e^{i \omega_0 t}$, yielding 
\begin{equation}
 i \dot{\phi}_{n}-\left[\omega_0 + V_0 (-1)^n\right] \phi_n + 
C (\phi_{n+1}+\phi_{n-1}) 
+ |\phi_{n}|^{2}\phi_{n} = 0 ,
 	\label{DNLSphi}
 \end{equation}
whereupon standard Newton-type schemes are used to find time-periodic 
solutions to Eq.\ (\ref{DNLSphi}) fulfilling 
$\phi_n(t+2\pi/\omega_b)= \phi_n(t)$, where $\omega_b$ is the second 
frequency. In addition, linear stability is simply investigated numerically 
by standard Floquet analysis as described in Ref.\ \cite{JA97}. 

When a 
stationary gap breather of frequency $\Lambda$ has an internal linear 
eigenmode of frequency $\omega_l$ (in the frame rotating with frequency 
$\Lambda$), 
the family of quasiperiodic solutions bifurcating 
from it at this point has $\omega_b = \omega_l$ and 
$\omega_0 = \Lambda - \omega_l$. Thus, by using as initial trial solution 
the stationary gap breather perturbed with the relevant linear eigenmode, 
the two-parameter family of solutions 
(for given $V_0$ and $C$) can then be followed through parameter space by 
standard continuation 
techniques. (Note that, in contrast to Ref.\ \cite{JA97}, once the
stationary gap breather is known we here obtain 
directly quasiperiodic solutions at finite $C$, without invoking the 
anticontinuous limit $C\rightarrow 0$.) For a small enough continuation 
step ($\lesssim 10^{-3}$), we typically with standard double 
precision 
fortran routines for numerical integration and matrix inversion obtain 
convergence to solutions with 
$||\phi_n(t+2\pi/\omega_b)- \phi_n(t)|| < 10^{-12}$ in five or less 
Newton 
iterations. 
In fact, since norm is a conserved quantity, it is generally 
more instructive to use ${\cal N}$ and $\omega_b$ as independent parameters, 
and practically continuations versus $\omega_b$ at constant norm may 
be 
performed by varying $\omega_0$ to keep  ${\cal N}$ constant. It is also 
convenient to work with rescaled quantities amounting to setting $C=1$ 
in Eqs. (\ref{DNLS}), (\ref{DNLSphi}); 
consequently the three relevant independent parameters for the two-frequency 
solutions can be chosen as 
$V_0/C$, ${\cal N}/C$ (or, alternatively, $\omega_0/C$), and $\omega_b/C$.

\begin{figure*}
\includegraphics[height=0.37\textwidth,angle=270]{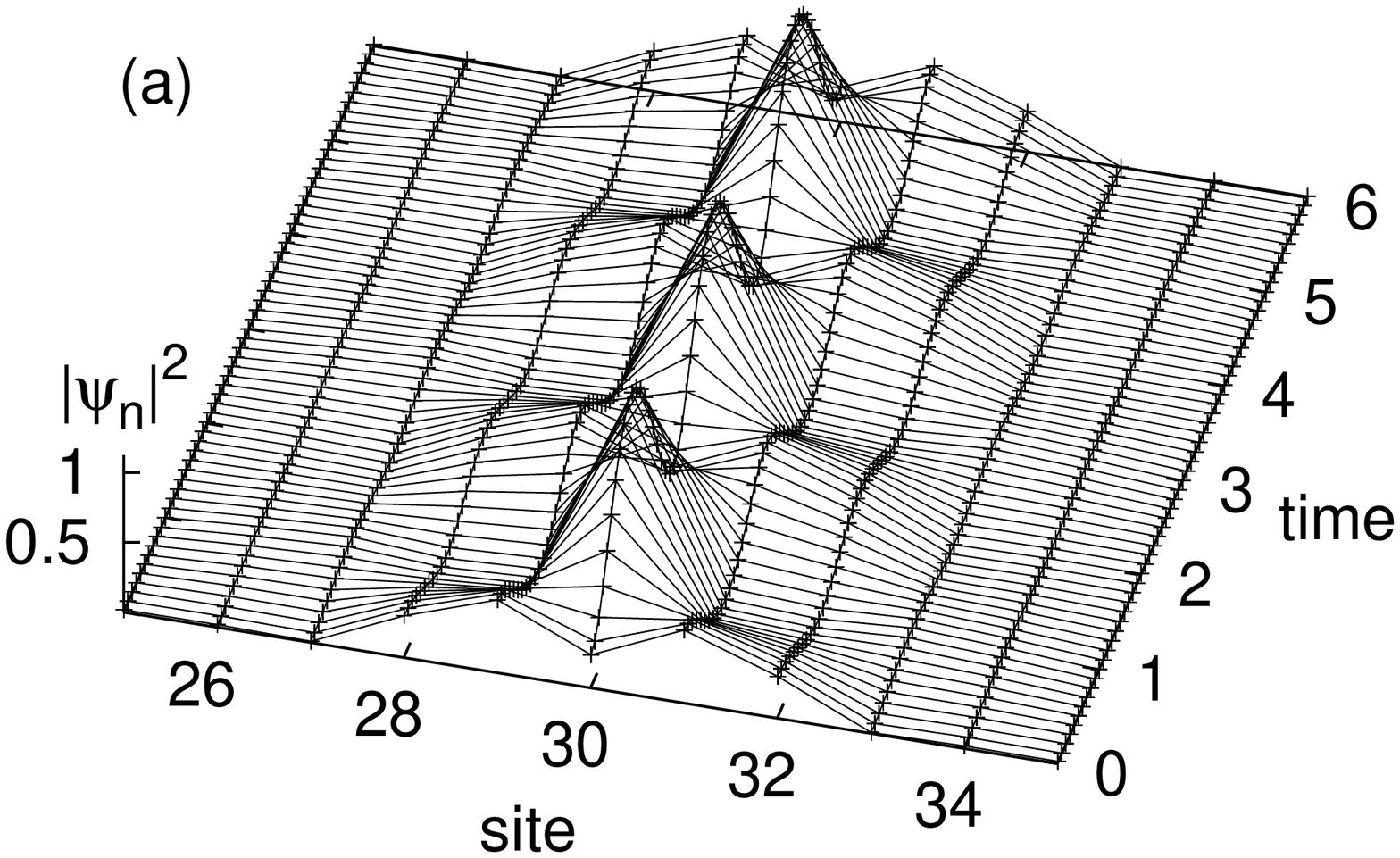}%
%\\
\includegraphics[height=0.2\textwidth,angle=270]{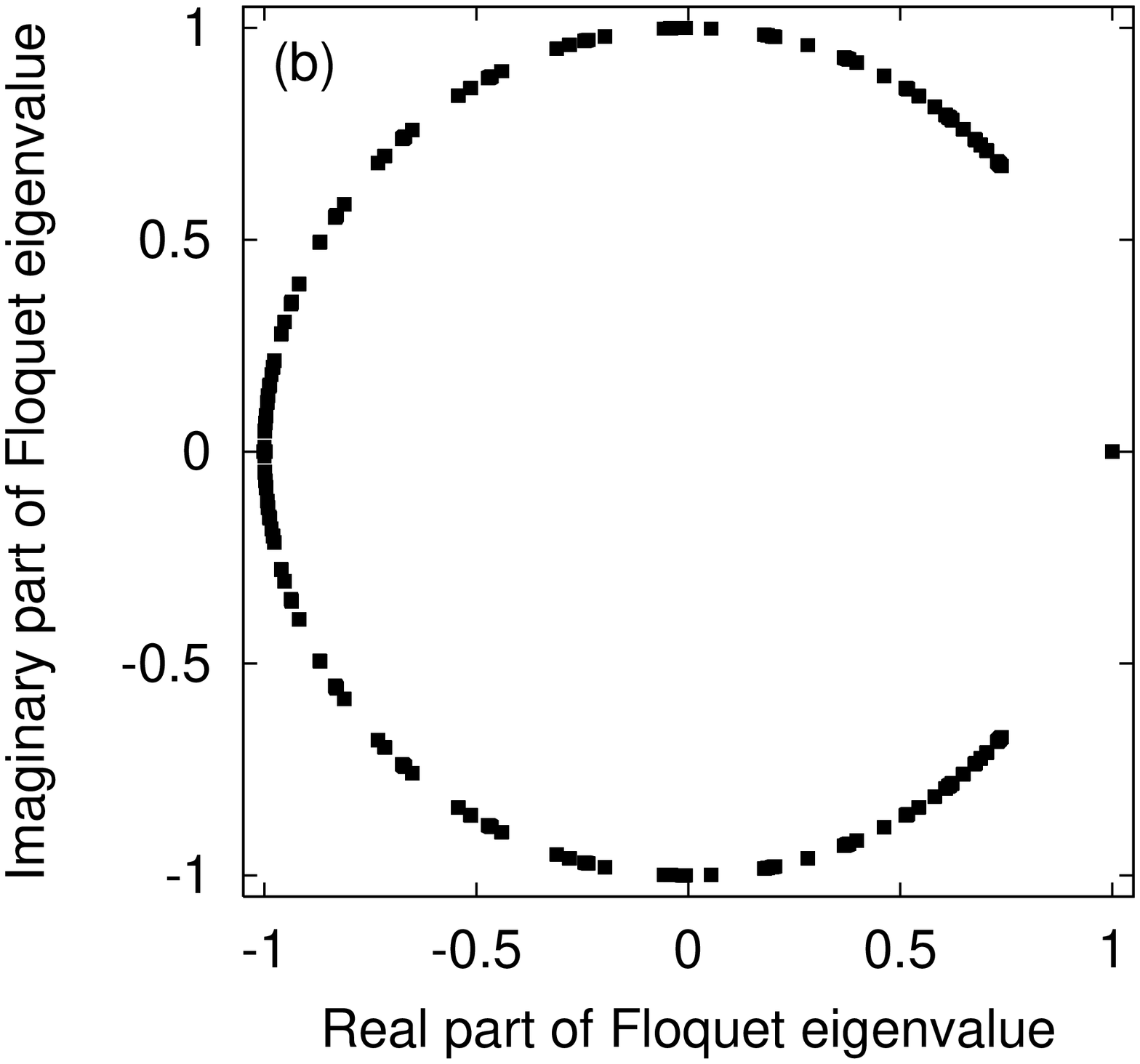}%
%\\
\includegraphics[height=0.37\textwidth,angle=270]{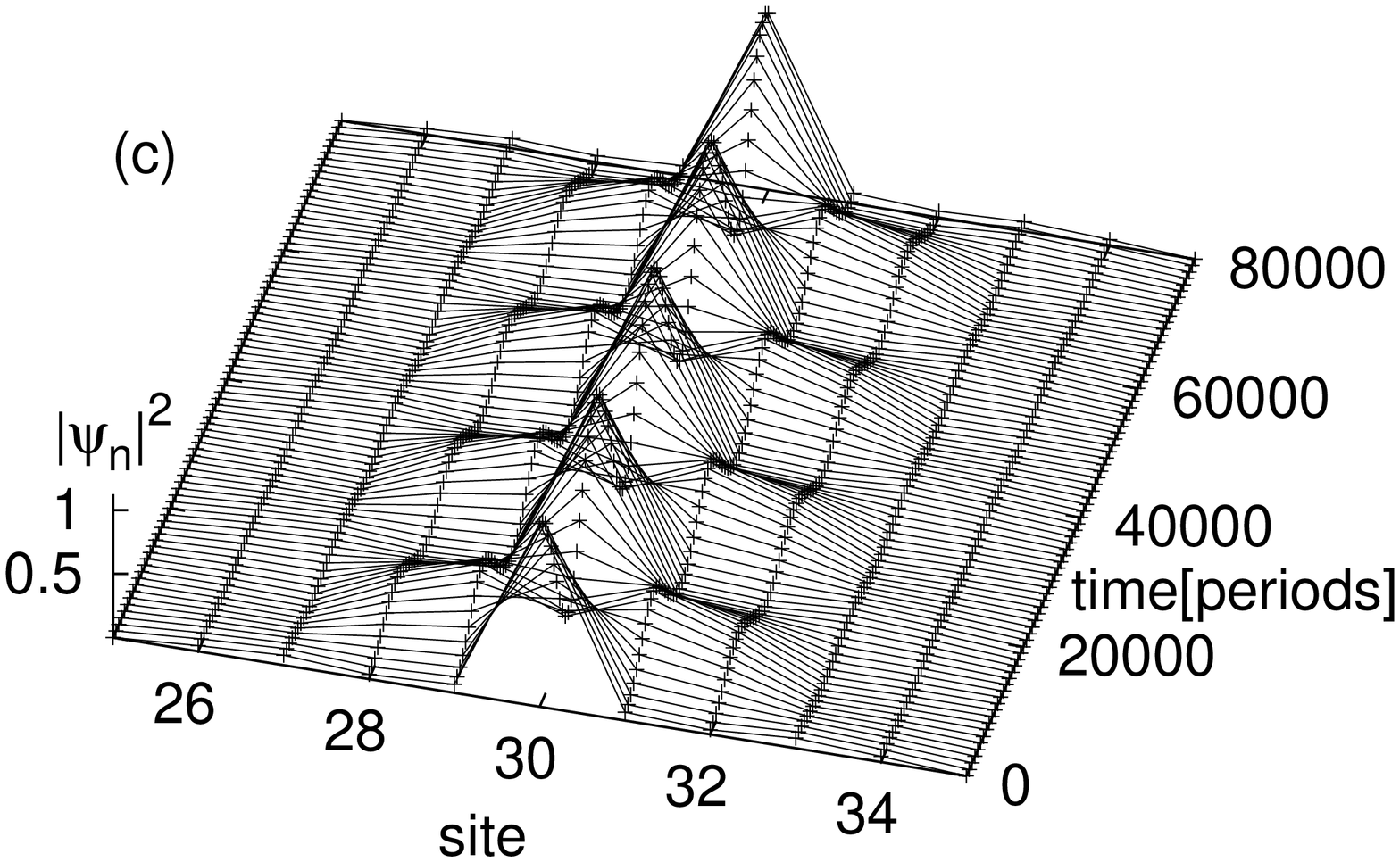}%
\caption{(a) The first three oscillations of an exact localized pulson 
solution 
to Eq.\ (\ref{DNLS}) with 
$V_0 = 1$, $C=1$. The dynamical parameters are $\omega_b=3.18$, 
$\omega_0 = -3.722$, ${\cal N}\approx 1.90867$. Only the central part is 
shown; the solution decays exponentially outside this region. (b) Linear 
stability eigenvalues for the solution in (a). 
(c) Snapshots, at integer multiples of the original period 
$2\pi / \omega_b$, from long-time integration using the solution in (a), 
with a random perturbation $\sim 10^{-3}$, as initial 
condition. }
\label{QP1}
\end{figure*}

Fig.\ \ref{QP1} illustrates the dynamics of a typical 
exact exponentially localized pulson solution, obtained by continuation 
(versus $\omega_b/C$ at constant $\omega_0/C$ and $V_0/C$) 
of the linear 'P1' mode of a stationary gap breather with $\Lambda=-0.5V_0$ 
(i.e., 
with stationary frequency in the middle of the lower half of the gap). 
The intensity oscillates periodically, from having a minimum at the central 
(even) site  and two symmetric maxima at the neighboring (odd) sites 
 at $t=0$, to having one single maximum at the central site 
at $t=\pi/\omega_b$. The solution is linearly stable, as is seen from 
the numerically calculated Floquet eigenvalues (Fig.\ \ref{QP1}(b)), 
which are all on the unit circle. The main effect of a small but 
non-negligible perturbation is illustrated by Fig.\ \ref{QP1} (c): the 
solution almost perfectly retains its pulson character over very large 
time scales but with a 
slight shift in frequency (a consequence of the four-fold degenerate 
Floquet eigenvalue at $+1$ corresponding to drifts of the two arbitrary 
phases \cite{JA97}), yielding visible the pulson oscillations also in 
the stroboscopic plot. 

\begin{figure}
\includegraphics[height=0.37\textwidth,angle=270]{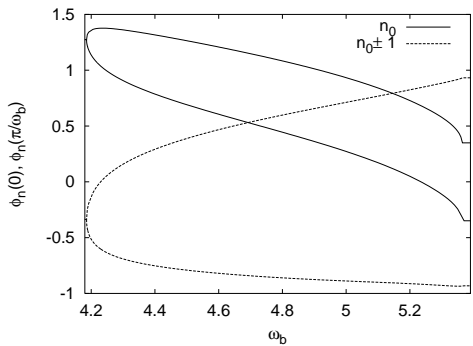}%
\caption{Maximum and minimum amplitudes for the central (even) site $n_0$ 
and its neighboring (odd) sites for continuation at constant norm 
${\cal N} = 2$ of the family of two-frequency solutions bifurcating from 
a stationary gap breather with  $\Lambda\approx -0.905$ at 
$\omega_b \approx 4.186$ (leftmost part of the figure). Here $V_0=2$ and 
$C=1$. The solution is unstable for $\omega_b\gtrsim 4.47$.}
\label{V2N2P1}
\end{figure}

\begin{figure}
\includegraphics[height=0.37\textwidth,angle=270]{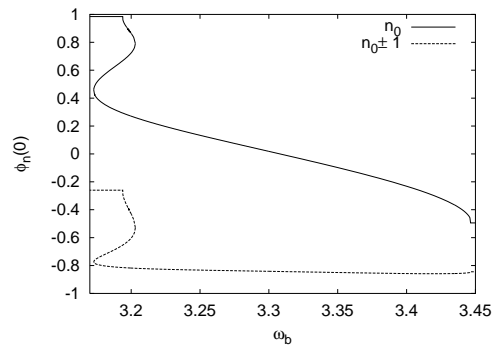}%
\caption{Same as Fig.\ \ref{V2N2P1} but for $V_0=1$ and 
with only minimum amplitudes. The two-frequency solution family bifurcates 
from a stationary gap breather with $\Lambda\approx -0.558$ at 
$\omega_b \approx 3.194$, and is unstable for $\omega_b\gtrsim 3.21$ 
on the lower branch.}
\label{V1N2P1}
\end{figure}

\begin{figure}
\includegraphics[height=0.37\textwidth,angle=270]{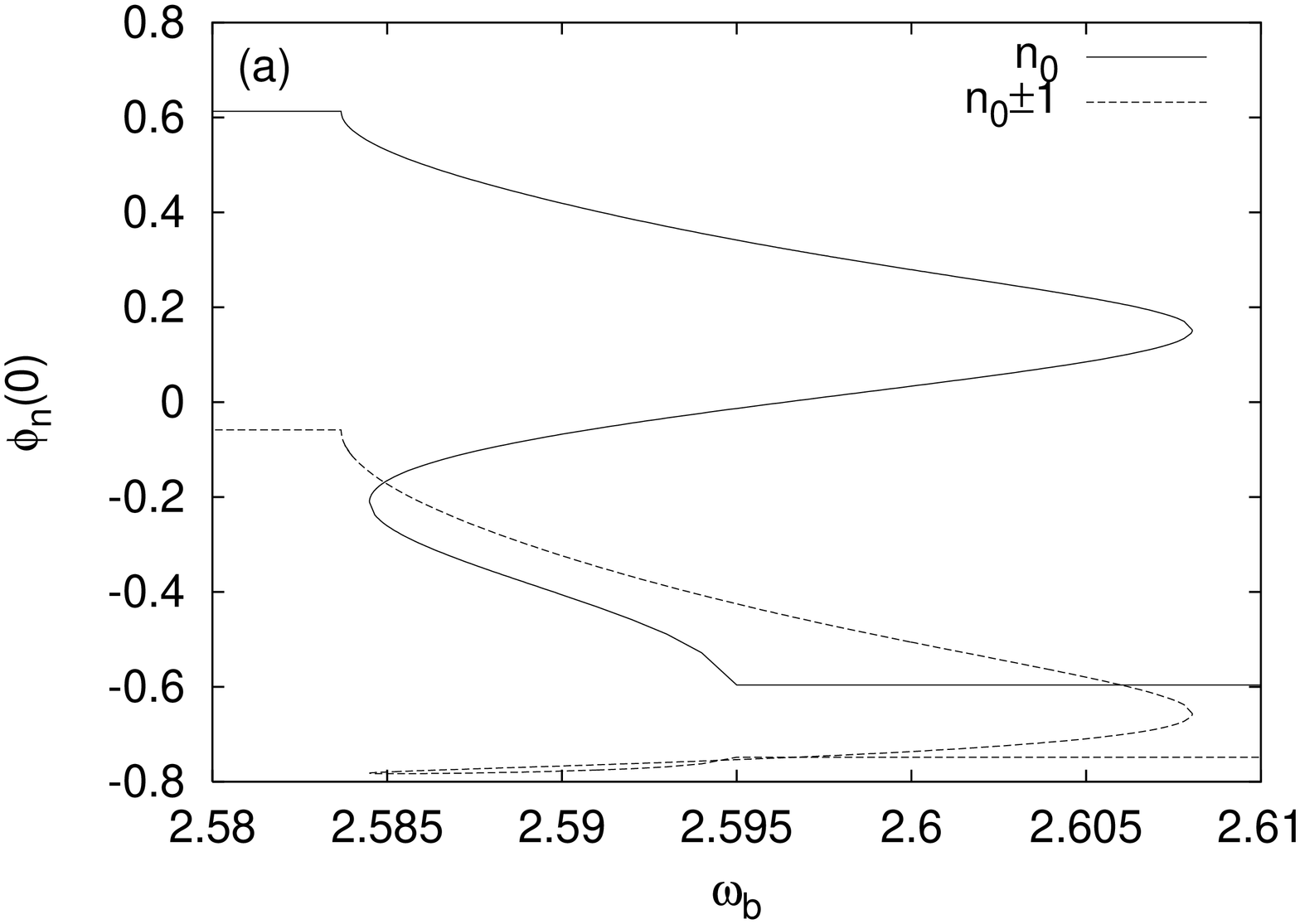}%
\\
\includegraphics[height=0.37\textwidth,angle=270]{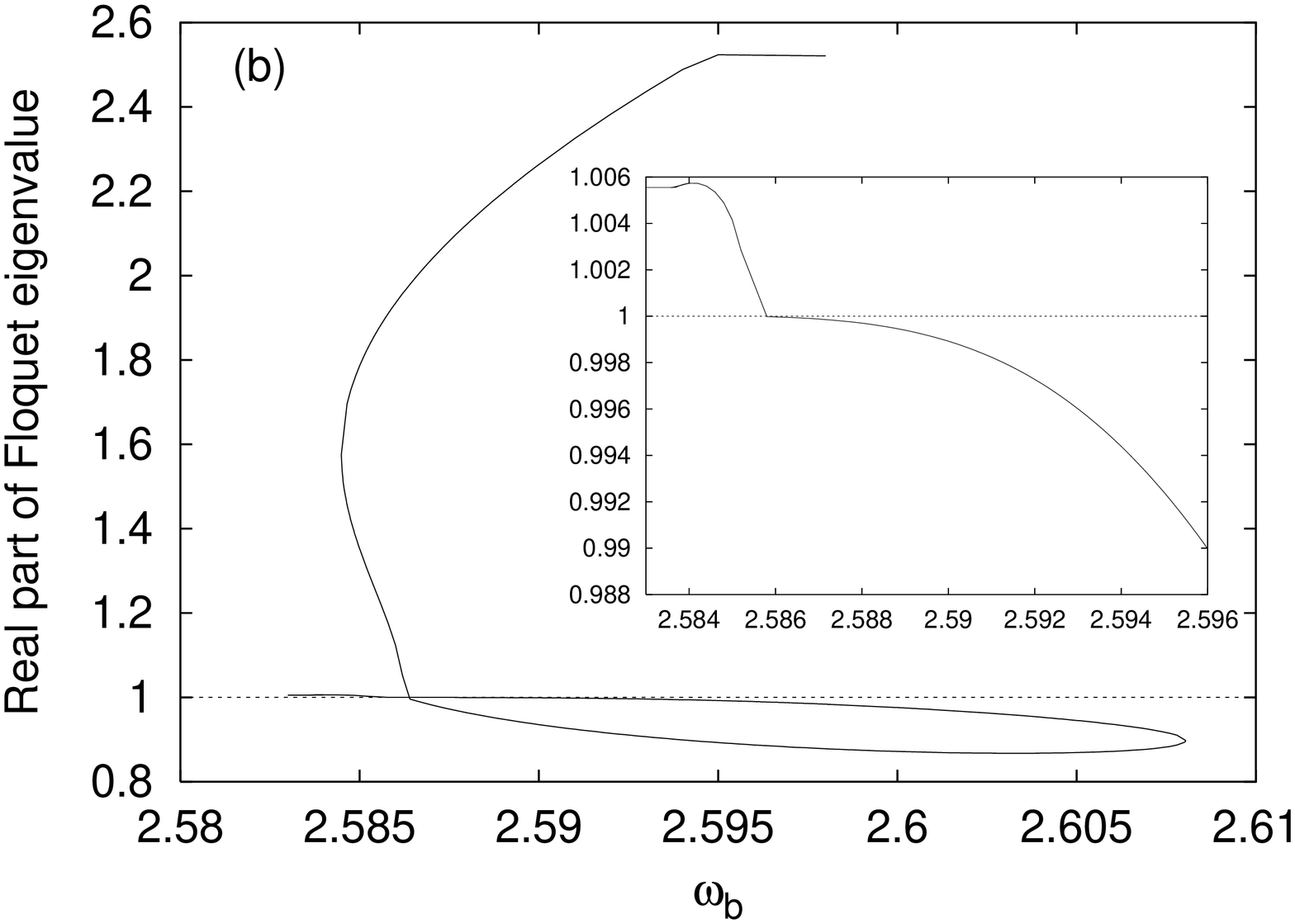}%
\caption{(a) Same as Fig.\ \ref{V1N2P1} but for $V_0=0.5$. 
The two-frequency solution family bifurcates 
from the stationary gap breather with $\Lambda\approx -0.3147$ at 
$\omega_b \approx 2.5837$, and (b) becomes unstable for 
$\omega_b\approx 2.5862$ on the middle branch. Magnification in inset 
of (b) illustrates how 
the weakly unstable stationary gap breather gets stabilized by the intensity 
oscillations. In (b), only the largest real part of the eigenvalue pair 
corresponding to the symmetry-breaking localized mode is shown.}
\label{V0.5N2P1}
\end{figure}

The results of more systematic investigations of the properties of such 
families of solutions are summarized in Figs.\ \ref{V2N2P1}-\ref{V0.5N2P1}. 
For 
convenience, we now choose a 'moderate' constant 
value of the norm, ${\cal N}/C=2$, 
and discuss the behaviour for various values of $V_0/C$. In particular, this 
value is large enough for the stationary gap breather frequency $\Lambda$
to be sufficiently far 
away from the lower gap boundary so to always have a localized 'P1' mode, 
but small enough for $\Lambda$ 
to be in the lower half of the gap, thus avoiding 
additional complications such as oscillatory instabilities and nonmonotonous 
continuation which may appear for stationary breathers 
in the upper gap half \cite{GJ03}.

Fig.\ \ref{V2N2P1}, for $V_0/C=2$, shows the typical behaviour for larger 
values of $V_0/C$. The continuation of the two-frequency solution 
is monotonous, and ends in a bifurcation with another (unstable) 
stationary solution 
with frequency $\Lambda=2 \omega_b+\omega_0$. With the notation of 
Ref.\ \cite{GJ03}, this solution is denoted as 
$\{\mathbb{O}\uparrow(\mathbb{O}0)\}_S^O$, meaning that, in the 
anticontinuous limit $C\rightarrow 0$, it corresponds to symmetric in-phase 
oscillations only for the two (odd) sites $n_0\pm1$, with frequency 
{\em above} 
the upper linear band. The pulson character of the two-frequency solution 
appears when the absolute value of the minimum (negative) value of 
$\phi_{n_0\pm1}$ (lower part of dashed curve in Fig.\ \ref{V2N2P1}) exceeds 
the minimum value of $\phi_{n_0}$ (lower part of solid curve), which 
for the case in Fig.\ \ref{V2N2P1} happens for 
$\omega_b\gtrsim 4.43$. However, a stable pulson solution appears in this 
case only in a small frequency interval, since for $\omega_b\gtrsim 4.47$ 
the solution is unstable through a symmetry-breaking instability (with real 
Floquet eigenvalue). For larger values of $V_0/C$, when the solution becomes 
more discrete, the solution becomes unstable before it attains its pulson 
character, and thus in the strongly discrete case the stable two-frequency 
solutions only correspond to relatively small oscillations of the 
central-site intensity for the stationary gap breather. 

For smaller values of $V_0/C$, the continuation versus $\omega_b$ 
of the 
two-frequency solutions at fixed norm ${\cal N}/C = 2$ still starts and 
ends in 
the same stationary solutions as before, 
but becomes nonmonotonous as shown for 
$V_0/C=1$ in Fig.\ \ref{V1N2P1} (for clarity only minimum amplitudes 
$\phi_n(0)$ are shown). The solution now attains its pulson character on the 
intermediate branch, where it is always stable, and becomes unstable 
only on the lower branch. 

For even smaller values of $V_0/C$, the stationary gap breathers 
broaden and assume a more 'continuous-like' shape, which in particular 
implies that the odd sites with largest $|\phi_n^{(\Lambda)}|$ 
will change from $n_0\pm1$ to  $n_0\pm3$, and further to $n_0\pm5$, 
etc.\ \cite{GJ03}.
For $V_0/C=0.5$ (Fig.\ \ref{V0.5N2P1}), the largest 
'odd' $|\phi_n^{(\Lambda)}|$ for the stationary breather is at $n_0\pm5$; 
however already for rather small intensity oscillations 
($\omega_b \gtrsim 2.5856$ on the upper branch of Fig.\ \ref{V0.5N2P1}(a)) 
of the corresponding 'P1' family of two-frequency solutions, the maximum 
'odd' intensity is again at $n_0\pm1$. 
The solution now attains its pulson 
character already on the upper branch ($\omega_b \gtrsim 2.5926$), and 
the pulson remains stable until it reaches 
$\omega_b\approx 2.5862$ on the middle branch (Fig.\ \ref{V0.5N2P1}(b)).

Another very interesting effect regarding stability is seen 
in Fig.\ \ref{V0.5N2P1}(b). It is known \cite{GJ03} that as $V_0/C$ 
decreases, there are certain intervals of 'inversion of stability' for 
stationary gap breathers, where the symmetric stationary breather discussed 
here becomes unstable (through a translational 'depinning' instability), 
and instead the 
otherwise unstable {\em antisymmetric} gap breather gains stability. 
$V_0/C = 0.5$ belongs to such an interval for ${\cal N}/C = 2$, and thus 
the stationary gap breather from which the two-frequency family in 
Fig.\ \ref{V0.5N2P1} bifurcates is itself unstable. However, as seen 
from the inset in Fig.\ \ref{V0.5N2P1}(b), already at 
$\omega_b \approx 2.5856$ the unstable eigenvalue returns to the unit 
circle, and thus even rather small symmetric intensity oscillations may 
{\em stabilize} the stationary solution with respect to its antisymmetric 
instability. We are not aware of any reported similar scenario. 

To conclude, we have presented explicit examples of exact stable 
quasiperiodic 
pulson solutions with large-amplitude intensity oscillations, in the binary 
modulated DNLS equation describing e.g.\ coupled waveguides of alternating 
widths. Thus, there should be good possibilities for experimental 
observation of such states in this context. Although we here focused on 
one particular family of symmetric solutions bifurcating from the symmetric 
stationary 
gap breather, the analysis illustrated by Fig.\ 3 in Ref.\ \cite{GJ03} 
suggests that solutions with similar properties may bifurcate from other, 
symmetric or antisymmetric, internal modes  with frequencies 
above the continuous spectrum 
('P2', 'P3', 'P4', 'P5', etc.), existing for the symmetric as well as the 
antisymmetric gap breathers. It seems likely, that such solutions also
should exist for other types of multicomponent lattices with (at least) two 
conserved quantities; one interesting candidate being
second-harmonic-generating 
lattices of the type considered e.g.\ in 
Ref.\ \cite{MKFNY02}, where also the 
fundamental discrete soliton was found to exhibit an internal mode with 
frequency above the linear spectrum. A challenge for future research 
is to obtain analytical expressions for the 
pulson solutions in the continuum limit. This is a nontrivial task, 
since although the relevant linear eigenmodes of the stationary gap breather 
apparently persist arbitrarily close to the continuum limit (cf.\ Fig.\ 3 
of Ref.\ \cite{GJ03}), they cannot exist in the standard continuous 
two-field model for gap solitons (which is the same for DNLS as for 
diatomic Klein-Gordon lattices, see e.g.\ Refs.\ \cite{KF92,KUG99}), 
since the continuous 
spectrum of such models extends to infinity leaving no room for 
localized modes above it. A more sophisticated continuous approximation 
would therefore appear necessary to this end.

Financial support from the Swedish 
Research Council is gratefully acknowledged.

% Create the reference section using BibTeX:
%\bibliography{basename of .bib file}

\begin{thebibliography}{99}

\bibitem{waveguides} 
D.N.\ Christodoulides and R.I.\ Joseph, Opt.\ Lett.\ {\bf 13}, 794 (1988); 
H.S.\ Eisenberg {\it et al.}, Phys.\ Rev.\ Lett.\ {\bf 81}, 3383 (1998).

\bibitem{BEC}
A.\ Trombettoni and A.\ Smerzi, Phys.\ Rev.\ Lett.\ {\bf 86}, 2353 (2001).


\bibitem{KRB01} 
P.G.\ Kevrekidis, K.\O.\ Rasmussen, and A.R.\ Bishop, 
Int. J. Mod. Phys. B {\bf 15}, 2833 (2001).

\bibitem{EJ} J.C.\ Eilbeck and M.\ Johansson, in 
\textit{Localization and Energy 
Transfer in Nonlinear Systems}, edited by L.\ V\'azquez,
R.\ S.\ MacKay, and M.P.\ Zorzano (World Scientific, Singapore, 2003), 
p.\ 44; arXiv: nlin.PS/0211049 (2002).

\bibitem{SK} A.A.\ Sukhorukov and Yu.S.\ Kivshar, Opt.\ Lett.\ 
\textbf{27}, 2112 (2002); \textbf{28}, 2345 (2003); 
Phys.\ Rev.\ Lett.\ \textbf{91}, 113902 (2003); 
A.A.\ Sukhorukov, Yu.S.\ Kivshar, H.S.\ Eisenberg, and
Y.\ Silberberg, IEEE J.\ Quantum Electron.\ \textbf{39}, 31 
(2003).

\bibitem{GJ03} 
A.V.\ Gorbach and M.\ Johansson, Eur.\ Phys.\ J.\ D {\bf 29}, 77 (2004).

\bibitem{gapexp}
R.\ Morandotti {\it et al.}, arXiv: physics/0405037 (2004).


\bibitem{MacKay}
R.\ S. MacKay and S.\ Aubry,
   Nonlinearity  {\bf 7}, 1623 (1994).

\bibitem{JA97} M.\ Johansson and S.\ Aubry, Nonlinearity {\bf 10}, 1151 
(1997); M.\ Johansson {\it et al.}, Physica D {\bf 119}, 115 (1998).

\bibitem{Nussbaum} I.\ Nussbaum, Phys.\ Lett.\ A {\bf 118}, 127 (1986).

\bibitem{BV02} D.\ Bambusi and D.\ Vella, Disc.\ Cont.\ Dyn.\ Syst. -B 
{\bf 2}, 389 (2002).

\bibitem{KW03} P.G.\ Kevrekidis and M.I.\ Weinstein, Math.\ Comput.\ 
Simul.\ {\bf 62}, 65 (2003).

\bibitem{Mez} V.K.\ Mezentsev, S.L.\ Musher, I.V.\ Ryzhenkova, and 
S.K.\ Turitsyn, JETP Lett.\ {\bf 60}, 829 (1994); P.L.\ Christiansen 
{\it et al.}, Phys.\ Rev.\ B {\bf 54}, 900  (1996).

\bibitem{growth}
M.\ Johansson and S.\ Aubry, Phys.\ Rev.\ E {\bf 61}, 5864 (2000). 

\bibitem{MKFNY02}
B.A.\ Malomed {\it et al.}, Phys.\ Rev. E {\bf 65}, 056606 (2002). 

\bibitem{KF92}
Yu.S.\ Kivshar and N.\ Flytzanis, Phys.\ Rev.\ A {\bf 46}, 7972 (1992).

\bibitem{KUG99}
A.S.\ Kovalev, O.V.\ Usatenko, and A.V.\ Gorbatch, Phys.\ Rev.\ E {\bf 60}, 
2309 (1999).

\end{thebibliography}

\end{document}